\documentclass[12pt,a4paper]{article}
\usepackage{amsmath}
\usepackage{amsthm}
\usepackage{amsfonts}
\usepackage{amssymb}
\usepackage{makeidx}
\usepackage[all]{xy}
\usepackage{verbatim}
\usepackage{xcolor}

\usepackage{natbib}

\usepackage{graphicx}

\begin{document}
\sloppy

\title{\textbf{Thermodynamics and SARS-CoV-2: neurological effects in post-Covid 19 syndrome}}
	
\author{Umberto Lucia $^{1,a}$, Giulia Grisolia $^{1,b}$ \& Thomas S. Deisboeck $^{2,c}$\\ $^1$ Dipartimento Energia ``Galileo Ferraris'', \\ Corso Duca degli Abruzzi 24, 10129 Torino, Italy\\ $^2$ Department of Radiology, \\ Harvard-MIT Martinos Center for Biomedical Imaging,\\ Massachusetts General Hospital and \\ Harvard Medical School, Charlestown, MA 02129, USA  \\ $^{a}$ umberto.lucia@polito.it \\  $^{b}$ giulia.grisolia@polito.it \\ $^c$ deisboec@helix.mgh.harvard.edu}

\date{}
\maketitle

\begin{abstract}
There is increasing evidence that infection with SARS-CoV-2 can cause a spectrum of neurological symptoms. In this paper, we develop a theoretical concept underlying such neurological COVID-19 consequences by employing a non-equilibrium thermodynamic approach that allows linking the neuronal electric potential with a virus-induced pH variation. Our theoretical findings support further experimental work on therapeutically correcting electrolyte imbalances, such as Na$^+$ and K$^+$, to attenuate the neurological effects of SARS-CoV-2.	

\textit{Keyword}: 
Irreversible thermodynamics; Membrane potential; Neurological effects; Post-Covid Syndrome; SARS-CoV-2.

\end{abstract}

\section{Introduction}
Infection with the new severe acute respiratory syndrome coronavirus 2, i.e., SARS-CoV-2, can lead to a variety of  clinical symptoms ranging from respiratory and circulatory effects to neurological ones; some of these symptoms require immediate therapeutic intervention to try to stabilize vulnerable patients, while others are feared to potentially cause long term morbidity in some.  \citep{guetal2020}. 

In this theoretical paper, we focus our analysis on the neurological consequences of SARS-CoV-2. Since the virus has seldom been found in the cerebrospinal fluid of the patients, the damage caused by the anti-virus immune response-mediated damage seems to be the culprit \citep{solomon2021cov}. Severe neurological manifestations, albeit rare by comparison, run the gamut from Guillain-Barre (demyelinating polyneuritis) syndrome, and ischemic stroke \citep{REF1-1,Ref1} to encephalitis \citep{elluletal2020} while milder symptoms include temporary memory loss, altered mental state or ‘brain fog’ as well as olfactory and gustatory dysfunctions \citep{REF2}. In fact, partial or complete loss of smell, dysosmia or anosmia, as well as dysgeusia (loss of taste) are common symptoms during a SARS-CoV-2 infection, even in the absence of any other symptoms \citep{solomon2021cov,nelly,kanberg,germany,mandolfi,jenny}; as to the loss of smell, the virus seems to infect the olfactory epithelium, rather than the sensory neurons themselves \citep{nelly,kanberg,germany,mandolfi,jenny}. This notwithstanding, in autopsy studies of patients dying of COVID viral RNA transcripts were found in brain tissue and viral proteins in the endothelial cells within the olfactory bulb \citep{rockefeller}; in another study, inflammatory changes appear particularly extensive in the brainstem in proximity to cranial nerve origins \citep{1474-4422}.

At present, the lasting consequences of this neuro-invasion are not fully understood, neither is the optimal therapeutic strategy as, for instance, targeting an over-active immune response with corticosteroids would be potentially dangerous in the presence of virus \citep{REF3}. In this paper, we therefore develop a non-equilibrium thermodynamic analysis of this mechanism, in order to suggest a possible new viewpoint that may hold promise for designing future therapies.

\section{Materials and Methods}
Protein phosphorylation is a fundamental biochemical mechanism regulating of the cell functions, due to its ability to activate and deactivate some enzymes and receptors \citep{rudolphetal2006,strong2002}. In this context the relation with kinases is of particular interest because kinases are related to cellular transduction signalling \citep{arditoetal2017}. 

Ions actively cross the cell membrane against its electrochemical potential by deriving the required energy from the hydrolysis of ATP, where the H$^+$-ATPase plays a fundamental role; this is related to the movement of positive charges into the cell, by generating large membrane voltage (inside negative and outside positive) and a pH gradient \citep{nakanishi-matsui,stevensetal1997,tuszynskietal2003}:
\begin{equation}\label{Eq15}
\text{ATP} + \text{H}_2\text{O} \rightarrow \text{ADP} + \text{P}
\end{equation}
\begin{equation}\label{Eq16}
\text{H}^+_{out} \rightarrow \text{H}^+_{in}
\end{equation}
with a subsequent variation of the pH because~\citep{tuszynski}:
\begin{equation}\label{Eq17}
\Delta \text{pH} = \frac{F}{2.3 RT}\,\big(\Delta\phi_m - \Delta G_{\text{H}^+}\big)
\end{equation}
where $G$ depicts the Gibbs potential. The phosphorylation potential, $\Delta\bar{g}_p$ [kJ mol$^{-1}$], is well known and described by the following equation \citep{tuszynskietal2003,grabeetal2000,lucia-biocells,lucia-bioeng-ijot,LG-romanianac,ULAPTSD2014scir,LGTSDAP-biochcarci,LG-seclawcell,LGMRA-Cleyes,LGetal-antchamtem,LG-cyano}:
\begin{equation}\label{eq-phosphorylationpotential}
\Delta\bar{g}_p=-nF\Delta\phi
\end{equation}
where $n$ is the number of moles of ions per ATP synthesized, $F = 96.485\times 10^3$ A s mol$^{-1}$ is the Faraday constant, and $\Delta\phi$ stands for the membrane potential. 

The movement of the ions can be analysed by introducing the Onsager general phenomenological relationships as they pertain to both the electrochemical potential and the heat flux \citep{yourgrauetal1982,callen,lucia-grisolia-life,lucia-grisolia-2020-Ca,lucia-grisolia-2020-glaucoma,goupil2011}:
\begin{equation}\label{eq3}
\left\{
\begin{array}{l}
\mathbf{J}_e =-L_{11}\,\dfrac{\nabla \mu_e}{T} - L_{12} \,\dfrac{\nabla T}{T^2} \\ \\
\mathbf{J}_Q =-L_{21}\,\dfrac{\nabla \mu_e}{T} - L_{22} \,\dfrac{\nabla T}{T^2}
\end{array}
\right.
\end{equation}
where $\textbf{J}_e$ is the current density [A m$^{-2}$], $\textbf{J}_Q$ denotes the heat flux [W m$^{-2}$], $\mu_e = \mu + ze\phi$ is the electrochemical potential [J mol$^{-1}$], with $\mu$ the chemical potential [J mol$^{-1}$], $ze$ the electric charge [A s mol$^{-1}$], and $\phi$ the membrane potential [V], respectively; $T$ is the living cell temperature and $L_{ij}$ represent the phenomenological coefficients, such that \citep{katchlskycurrant} $L_{12}(\mathbf{B}) = L_{21}(-\mathbf{B})$ (Onsager-Casimir relation \citep{degrootmazur}), and $L_{11} \geq 0$ and $L_{22}\geq 0$, and \citep{katchlskycurrant} $L_{11}L_{22}-L_{12}L_{21}>0$. 

The result consists of a model of the life cycle based on two related processes \citep{lucia-grisolia-life,lucia-grisolia-2020-Ca}:
\begin{itemize}
	\item A continuous energy generation (metabolism), due to ion fluxes: The ion and metabolite fluxes can be described by imposing $\mathbf{J}_e \neq \mathbf{0}$ and $\mathbf{J}_Q = \mathbf{0}$;
	\item A continuous heat flux from the cell to its microenvironment: The heat exchange towards the environment can be described by imposing $\mathbf{J}_e = \mathbf{0}$ and $\mathbf{J}_Q \neq \mathbf{0}$.
\end{itemize}
In this way, we can split the life cycle into two thermodynamic processes, as it is usually done in irreversible thermodynamics for any complex process \citep{callen}.

Now, if ion and metabolite fluxes occur, $\mathbf{J}_e \neq \mathbf{0}$ and $\mathbf{J}_Q = \mathbf{0}$, it follows \citep{callen,yourgrauetal1982,lucia-grisolia-life}
\begin{equation}\label{eq3flux}
\frac{d\mu_e}{dT} = -\frac{L_{21}}{L_{11}}\,\frac{1}{T}
\end{equation}
with a related heat flux \citep{callen,yourgrauetal1982}:
\begin{equation}\label{equJ}
\frac{du}{dt} = -\nabla\cdot\mathbf{J}_Q
\end{equation}
where $u$ is the internal energy density [W m$^{-3}$]. 

Living cells exchange heat power towards their environment by convection, and so, we can write~\citep{lucia-grisolia-2020-cancer-entropy}
\begin{equation}\label{eq7}
\frac{du}{dt}\, dV = \delta\dot{Q} =  - \alpha\,(T-T_0)\,dA 
\end{equation}
where \mbox{$\alpha \approx 0.023 Re^{0.8} Pr^{0.35} \lambda/\langle R\rangle$} is the coefficient of convection, $A$ the area of the external surface of the cell membrane, $V$ is the cell volume, $T$ depicts the mean temperature of the external surface of the cell's membrane, and $T_0$ is the temperature of the cell environment. 

So, considering Equations (\ref{equJ}) and (\ref{eq7}), and the Divergence Theorem \citep{apostol}, the heat flux can be written as:
\begin{equation}\label{eq6jq}
J_Q=  \alpha\,(T-T_0)
\end{equation}
and the related power flux yields:
\begin{equation}\label{eqQQ}
\dot{Q} = \int_A\mathbf{J}_Q \cdot \hat{\mathbf{n}}dA =  \alpha\,(T-T_0) A 
\end{equation}

Furthermore, considering Equation~(\ref{eq3}), together with the second hypothesis of our modelling  ($\mathbf{J}_e=\mathbf{0}$, $\mathbf{J}_Q\neq \mathbf{0}$), it follows \citep{lucia-grisolia-life}:
\begin{equation}\label{eq8}
\frac{d\mu_e}{d\ell} = \frac{T\, J_Q}{\Bigg(L_{22}\frac{L_{11}}{L_{12}}-L_{21} \Bigg)} =- \frac{\alpha\,T(T-T_0)}{\Bigg(L_{22}\frac{L_{11}}{L_{12}}-L_{21} \Bigg)}
\end{equation}
where $\ell$ is the length of a cell membrane and $|\nabla\mu_e| \approx d\mu_e/d\ell$. This relation is the link between the cell membrane's electric potential and the temperature of the cell itself.

Equations (\ref{eq6jq}) and (\ref{eq8}) allow us to obtain:
\begin{equation}
J_Q = \alpha\,(T-T_0) = - \frac{1}{T}\Bigg(L_{22}\frac{L_{11}}{L_{12}}-L_{21} \Bigg)\, \frac{d\mu_e}{d\ell}
\end{equation}
where:
\begin{equation}\label{eq11}
\Bigg(L_{22}-L_{21} \frac{L_{12}}{L_{11}}\Bigg) = K_J T^2
\end{equation}
with $K_J$ being the Thomson coefficient. Consequently, it follows:
\begin{equation}\label{eq12a}
\frac{\partial \mu_e}{\partial \ell} =  \frac{\partial \mu_e}{\partial T} \, \frac{\alpha}{K_J} \big(T_{surf}-T_0\big)
\end{equation}
from which, taking into account that $\mu_e =\mu+ze\phi$, becomes:
\begin{equation}\label{eq12b}
\frac{\partial \mu}{\partial \ell} = - ze\frac{d\phi}{d\ell} + \frac{\partial \mu_e}{\partial T} \, \frac{\alpha}{K_J} \big(T_{surf}-T_0\big)
\end{equation}
Now, considering Equation (\ref{Eq17}) we can obtain:
\begin{equation}\label{eq226}
\frac{\partial \mu_e}{\partial T} =\frac{K_J}{\alpha} \frac{F+ ze}{T_{surf}-T_0}\,\frac{d \phi}{d \ell} -  \frac{K_J}{\alpha} \frac{2.3R\,T_0}{T_{surf}-T_0} \,\frac{d\text{pH}}{d \ell} 
\end{equation}
which links the electrochemical potential to the pH.

In order to understand the effect of SARS-CoV-2 on the brain, we use a simple model of information coding, introduced previously in our thermodynamic analysis of Alzheimer’s disease \citep{lucia-grisolia-alz}. 

Specifically, a brain cell requires a Na$^+$-inflow, and a countering flow of K$^+$-outflow to develop the functionality of processing signals \citep{17}. During this function, the consumption of one molecule of ATP (adenosine triphosphate) requires that the membrane pump extrudes 3 Na$^+$-ions and imports 1 K$^+$-ion. At the stationary state, a neuron maintains its pump~current:
\begin{equation}\label{Eq1}
I_p = \frac{\Delta\phi_{\text{Na}^+} - \Delta\phi_m}{R_{\text{Na}^+}} + \frac{\Delta\phi_{\text{K}^+} - \Delta\phi_m}{R_{\text{K}^+}}
\end{equation}
where $R_i$ ($i=$ [Na$^+$] or [K$^+$]) stands for the electric resistance of the ion considered during its current flux through the membrane, and $\Delta\phi_m$ is the membrane electric potential variation, $\Delta\phi_{\text{Na}^+}$ is the electric potential variation due to the Na$^+$-flux \citep{20,goldman1943,tuszynski}:
\begin{equation}\label{Eq99}
\Delta\phi_{\text{Na}^+} = -\frac{RT}{F}\,\ln \Bigg(\frac{[\text{Na}^+]_f}{[\text{Na}^+]_i}\Bigg)
\end{equation}
$\Delta\phi_{\text{K}^+}$ is the electric potential variation due to the K$^+$-flux \citep{20,goldman1943,tuszynski}:
\begin{equation}\label{Eq1010}
\Delta\phi_{\text{K}^+} = -\frac{RT}{F}\,\ln \Bigg(\frac{[\text{K}^+]_f}{[\text{K}^+]_i}\Bigg)
\end{equation}
where $R = 8314$ J mol$^{-1}$ K$^{-1}$ denotes the constant of the ideal gasses, $F = 96,485$ C mol$^{-1}$ is the  Faraday constant, $f$ and $i$ means final and initial respectively, and they are referred to the initial and finale state of the neuronal signalling process, and~$T$ is the temperature, and
\begin{equation}
R_{in}=\frac{1}{\dfrac{1}{R_{\text{Na}^+}}+\dfrac{1}{R_{\text{K}^+}}}
\end{equation}
under the biochemical constraint:
\begin{equation}\label{Eq223}
\frac{d [\text{Na}^+]}{dt}=-\frac{d [\text{K}^+]}{dt}
\end{equation}
where [A] is the concentration of the A-ion (Na$^+$/K$^+$). In order to maintain a normal membrane potential of around~\citep{20} -70 mV  a neuron ($R_{in} =$ 200 M$\Omega$ of input resistance) requires an influx of around $1.02\times 10^9$ Na$^+$-K$^+$ ions s$^{-1}$ ($\Delta\phi_{Na^+} = - 50$ mV and $\Delta\phi_{K^+} = - 100$ mV) which necessitates $3.42 \times 10^8$ hydrolysed ATP molecules s$^{-1}$~\citep{20}, consumed at a rate of $I_p/F$: it generates a pump current $I_p$ of $1.63\times 10^{-10}$ A. 

Next, considering Equations (\ref{Eq223}), (\ref{Eq99})  and (\ref{Eq1010}) it follows that (\ref{Eq1010}) becomes:
\begin{equation}\begin{aligned}
&\frac{\partial \mu_{e,\text{Na}^+}}{\partial T} =-\frac{K_JRT_0}{F\alpha} \frac{F+ ze}{T_{surf}-T_0}\,\frac{1}{[\text{Na}^+]}\,\frac{d [\text{Na}^+]}{d \ell} -  \frac{K_J}{\alpha} \frac{2.3R\,T_0}{T_{surf}-T_0} \,\frac{d\text{pH}}{d \ell} \\
&\frac{\partial \mu_{e,\text{K}^+}}{\partial T} =-\frac{K_JRT_0}{F\alpha} \frac{F+ ze}{T_{surf}-T_0}\,\frac{1}{[\text{K}^+]}\,\frac{d [\text{K}^+]}{d \ell} -  \frac{K_J}{\alpha} \frac{2.3R\,T_0}{T_{surf}-T_0} \,\frac{d\text{pH}}{d \ell}
\end{aligned}
\end{equation}
which points out that, in order to maintain a normal chemical potential the cell, or neuron in this case, must actively change its concentration of ions if a change in the pH occurs as a result of the viral infection.

As a consequence of the previous steps, a density entropy rate due to irreversibility (dissipation function \citep{yourgrauetal1982}) is generated~\citep{LG-seclawcell}:
\begin{equation}\label{eq88}
\sigma =- \frac{1}{T_0}\, \sum_{i=1}^N\mathbf{J}_i\cdot\nabla\mu_i \geq 0
\end{equation}
where  $T_0$ represents the environmental temperature, $\sum_{i=1}^N\mu_i\,\mathbf{J}_i$ is the contribution of the inflows and outflows,  and~$\mu$ denotes the chemical potential. So, using the previous relations we obtain:
\begin{equation}\begin{aligned}
\sigma &\approx  
-\frac{R\, J_{\text{Na$^+$}}\,(ze)_{\text{Na$^+$}}}{F\ell}\,\ln \Bigg(\frac{[\text{Na}^+]_f}{[\text{Na}^+]_i}\Bigg) + \frac{J_{\text{Na$^+$}}}{T_0}\,\frac{\partial \mu_{e,\text{Na$^+$}}}{\partial T} \, \frac{\alpha}{K_J} \big(T_{surf}-T_0\big)
+ \\
&- \frac{R\, J_{\text{K$^+$}}\,(ze)_{\text{K$^+$}}}{F\ell}\,\ln \Bigg(\frac{[\text{K}^+]_f}{[\text{K}^+]_i}\Bigg) + \frac{J_{\text{K$^+$}}}{T_0}\,\frac{\partial \mu_{e,\text{K$^+$}}}{\partial T} \, \frac{\alpha}{K_J} \big(T_{surf}-T_0\big)
\end{aligned}
\end{equation}

Starting from this last equation, and considering the previous condition of non-negative entropy density \citep{katchlskycurrant}, we arrive at the following condition:
\begin{equation}\begin{aligned}
\frac{R\, J_{\text{Na$^+$}}\,(ze)_{\text{Na$^+$}}}{F\ell}\,&\ln \Bigg(\frac{[\text{Na}^+]_f}{[\text{Na}^+]_i}\Bigg) - \frac{J_{\text{Na$^+$}}}{T_0}\,\frac{\partial \mu_{e,\text{Na$^+$}}}{\partial T} \, \frac{\alpha}{K_J} \big(T_{surf}-T_0\big)
\leq  \\
&- \frac{R\, J_{\text{K$^+$}}\,(ze)_{\text{K$^+$}}}{F\ell}\,\ln \Bigg(\frac{[\text{K}^+]_f}{[\text{K}^+]_i}\Bigg) + \frac{J_{\text{K$^+$}}}{T_0}\,\frac{\partial \mu_{e,\text{K$^+$}}}{\partial T} \, \frac{\alpha}{K_J} \big(T_{surf}-T_0\big)
\end{aligned}
\end{equation}
which suggests that to maintain stability of the neuronal cell system the effect of sodium fluxes is less pronounced than that of potassium.

\section{Results}
Our conjecture yielded  Equations (\ref{Eq17}), (\ref{eq12b}) and (\ref{Eq1}). These equations highlight the link between the neuronal signalling process and the neurons’ membrane transport. 

Specifically, Equation (\ref{Eq17}) states that a change in pH determines a related variation in membrane potential and in proton flux, related to Gibbs energy and chemical potential (Equation (\ref{eq12b})). As such, a variation in the neuronal ion current pump occurs, with the consequence of modifying the membrane potential related to Na$^+$ and K$^+$. In turn, this change determines a modulation in the concentrations of these chemical species with a symmetry breaking in the stationary condition for the neurons. Consequently, the signalling process changes which may offer an explanation for the neurological consequences of infection with SARS-CoV-2. Indeed, we proved \citep{sars-cov-2-ns} that SARS-CoV-2 leads to changes in pH-homeostasis due to modifications of H$^+$-fluxes. Taken together, this suggests that a promising therapeutic strategy would seek to control the neurons’ membrane potential through manipulation of the ions responsible for it.

\section{Discussion and Conclusions}
In patients COVID-19 is characterized by a wide variety of symptoms, some of them neurological with at times very significant morbidity \citep{REF3}. To shed more light onto this, we have developed our non-equilibrium thermodynamics approach that focuses on the effect of SARS-CoV-2-induced pH variation on the ion and thermal fluxes across the cell membrane. As emphasized in Eq. 16 and 22, this `pH-ion flux' link can potentially support the development of new therapeutic strategies to combat SARS-CoV-2's neuronal effects. 

Electrolyte imbalances are common in Covid-19 patients, including hypokalemia which has been found to be an independent predictor of the requirement for mechanical ventilation \citep{S1201971220307499} and which may predispose to cardiac complications and necessitate potassium supplement therapy \citep{10.1001/jamanetworkopen.2020.11122}, and hypocalcemia  \citep{227080} which can lead to muscle twitches and tremors.  Furthermore, hyponatremia i.e. a serum sodium [Na$^+$] level of below 135 mmol L$^{-1}$ can cause serious central nervous system symptoms ranging from headaches, lethargy, and cramps to seizures, coma, and respiratory arrest \citep{add-ref-1}. Interestingly, hyponatremia has indeed been noted as an early sign in Covid-19 infections \citep{add-ref-2}. Clinical management of this common, multifactorial sodium imbalance in Covid-19 patients depends on the exact etiology and therefore can involve electrolyte replacement therapy in patients suffering from hyponatremia primarily due to fluid losses or infusion of hypertonic saline in cases of insufficient antidiuretic hormone secretion, seen as a consequence of the systemic inflammation triggered by the virus \citep{add-ref-2}.  It is intriguing in this context that a recent article, available so far only as preprint, describes in vitro experiments where hypertonic saline (1.5\% NaCl) inhibits SARS-Cov-2 replication completely, presumed to be achieved by cell plasma membrane depolarization and intracellular energy deprivation  \citep{preprint}. If these experimental findings can be confirmed, our  theoretical conjecture supports advancing this concept from in vitro to in vivo studies, in an effort to gain insights if hypertonic saline could help mitigate SARS-CoV-2 effects on the central nervous system. Finally, in light of these very recent experimental findings on sodium, Eq. 25 may deserve attention as it indicates that the impact of correcting potassium levels may be even more pronounced. Still, it must be noted that in clinics addressing electrolyte imbalances is highly non-trivial and extensive, experimental work would be required to properly assess risk vs. benefit of serum adjustments of Na$^+$ and/or K$^+$.  

In summary, our thermodynamics approach conceptualizes that SARS-CoV-2-induced pH fluctuations trigger ion-flux changes across the neuron cell membrane which in turn alters signalling throughout the system. Cautiously extrapolated, this may explain some of the neurological symptoms seen in COVID patients and it supports further research in therapeutically addressing electrolyte imbalances.

\section*{Authors contributions}
Conceptualization, U.L and T.S.D.; methodology, U.L., G.G. and T.S.D.; software, G.G.; validation, U.L., T.S.D. and G.G.; formal analysis, U.L.; investigation, G.G.; resources, U.L.; data curation, G.G.; writing-original draft preparation, U.L., G.G. and T.S.D.; writing—review and editing, U.L., G.G. and T.S.D.; visualization, G.G.; supervision, U.L., T.S.D.; project administration, U.L.; funding acquisition, U.L. All authors have read and agreed to the published version of the manuscript.

\bibliographystyle{apa-good}
\bibliography{document}

\end{document}